\documentclass[twocolumn]{aastex631}

\shorttitle{Elemental Abundance Diagnostics}

\shortauthors{Brooks et al.}
\journalinfo{To be submitted to The Astrophysical Journal}

\usepackage{grffile}

\begin{document}

\title{An elemental abundance diagnostic for coordinated Solar Orbiter/SPICE and Hinode/EIS observations}

\author[0000-0002-2189-9313]{David H. Brooks$^*$}
\affil{Computational Physics, Inc., Springfield, VA 22151, USA}
\affil{Department of Physics \& Astronomy, George Mason University, 4400 University Drive, Fairfax, VA 22030, USA}
\affil{University College London, Mullard Space Science Laboratory, Holmbury St. Mary, Dorking, Surrey, RH5 6NT, UK}
\altaffiliation{Current address: Hinode Group, ISAS/JAXA, 3-1-1 Yoshinodai, Chuo-ku, Sagamihara, Kanagawa 252-5210, Japan}

\author[0000-0001-6102-6851]{Harry P. Warren}
\affil{Space Science Division, Naval Research Laboratory, Washington, DC 20375, USA}

\author[0000-0002-0665-2355]{Deborah Baker}
\affil{University College London, Mullard Space Science Laboratory, Holmbury St. Mary, Dorking, Surrey, RH5 6NT, UK}

\author[0000-0001-9346-8179]{Sarah A. Matthews}
\affil{University College London, Mullard Space Science Laboratory, Holmbury St. Mary, Dorking, Surrey, RH5 6NT, UK}

\author[0000-0003-2802-4381]{Stephanie L. Yardley}
\affil{Department of Mathematics, Physics \& Electrical Engineering, Northumbria University, Ellison Place, Newcastle upon Tyne, NE1 8ST, UK}
\affil{University College London, Mullard Space Science Laboratory, Holmbury St. Mary, Dorking, Surrey, RH5 6NT, UK}
\affil{Donostia International Physics Center, Paseo Manuel de Lardizabal 4, San Sebasti\'an, 20018, Spain}

\begin{abstract}
Plasma composition measurements are a vital tool for the success of current and future solar missions, but density
and temperature insensitive spectroscopic diagnostic ratios are sparse, and
their underlying accuracy in determining
the magnitude of the First Ionization Potential (FIP) effect in the solar atmosphere remains an open question.
Here we assess the \ion{Fe}{8} 185.213\,\AA/\ion{Ne}{8} 770.428\,\AA\, intensity ratio that can be observed as a multi-spacecraft combination between
Solar Orbiter/SPICE and Hinode/EIS. We find that it is fairly insensitive
to temperature and density in the range of $\log$ (T/K) = 5.65-6.05 and is therefore useful, in principle, for analyzing on-orbit EUV spectra.
We also perform an empirical experiment, using
Hinode/EIS measurements of coronal fan loop temperature distributions weighted by randomnly generated FIP bias values, to
show that our diagnostic method can provide accurate results as it recovers the input FIP bias to within 10--14\%. This is encouraging since it is smaller than the magnitude of variations
seen throughout the solar corona. We apply the diagnostic to coordinated observations from 2023 March,
and show that the combination of SPICE and EIS allows measurements of the Fe/Ne FIP bias in the regions where the footpoints of the
magnetic field connected to Solar Orbiter are predicted to be located. The results show an increase in FIP bias between the main leading polarity and the
trailing decayed polarity that broadly
agrees with Fe/O in-situ measurements from Solar Orbiter/SWA. Multi-spacecraft coordinated observations are complex, but this diagnostic also falls within the planned wavebands
for Solar-C/EUVST.
\end{abstract}

\section{Introduction}
Plasma diagnostics underpin attempts to understand the processes that form the solar atmosphere and solar wind, and create the instabilities
that drive jets, flares, and coronal mass ejections. Elemental abundance measurements are a key component of these studies since they have
significant diagnostic value as a result of spatial and temporal variations due to the first ionization potential (FIP) effect. 

Early work
by \cite{Pottasch1963} suggested that elements with a low-FIP ($<$10\,eV) are enhanced in the solar corona by factors of 2--4 compared to their
photospheric abundances. This effect was later confirmed in the solar wind and solar energetic particles \citep[SEPs,][]{Meyer1985}. 
The first ionization energies for all the elements discussed in this paper are provided in Table \ref{table}.

Nowadays studies of the FIP effect and associated abundance variations
touch on all aspects of solar coronal physics. 
They have been used as a tracer of the source locations of the solar wind: establishing a connection between active
region outflows \citep{Sakao2007} and the slow solar wind \citep{Brooks2011,Brooks2015}, and verifying the expected photospheric composition of fast wind
sources such as polar jets \citep{Lee2015}. They have been used to study the basic composition structure of active regions: showing that the
hot cores have abundances enhanced by a factor of $\sim$ 3 \citep{DelZanna2014}, and that the bright fan loops and high temperature outflows at the
edges of active regions show a strong FIP Effect \citep{Warren2016,Brooks2011}. Much progress has also been made investigating the evolution of 
active region plasma composition, showing that the interaction with the preexisiting magnetic environment during the emergence phase,
and the surrounding quiet Sun in the decay phase, has an important influence \citep{Baker2013,Baker2015,Mihailescu2022}. 

Flares and eruptions \citep{James2017,Baker2021} have also been a target of abundance studies. Sun-as-a-star flare observations suggest that the composition is basically
photospheric - a response to evaporation after energy deposition in the chromosphere \citep{Warren2014}, but spatial variations have been found
that suggest the situation is more complex \citep{Doschek2018,To2021}.
Plasma emission from the presumed current sheet that
formed above an X8.3 flare arcade showed coronal composition \citep{Warren2018}, and
indeed sometimes an inverse FIP (IFIP) effect, where low-FIP elements are 
depleted or high-FIP elements are enhanced, is even seen 
in localized patches near strong magnetic field \citep{Doschek2015,Doschek2017,Baker2019,Baker2020}. The IFIP effect is well-known on active stars such as M-dwarfs \citep{Testa2015},
but was only recently discovered on the Sun \citep{Doschek2015} and in the slow solar wind \citep{Brooks2022}. 

\begin{deluxetable}{lcc}
\tabletypesize{\small}
\tablecaption{Ionization Energy Data}
\tablehead{
\multicolumn{1}{l}{Element} &
\multicolumn{1}{c}{Symbol} &
\multicolumn{1}{c}{Ionization Energy [eV]} 
}
\startdata
Calcium & Ca & 6.113 \\
Nickel & Ni & 7.640 \\
Magnesium & Mg & 7.646 \\
Iron & Fe & 7.902 \\
Silicon & Si & 8.152 \\
Sulfur & S & 10.360 \\
Oxygen & O & 13.618 \\
Argon & Ar & 15.760 \\
Neon & Ne & 21.565 
\enddata
\tablenotetext{}{Data from the NIST atomic database \citep{Kramida2023}.}
\label{table}
\end{deluxetable}

Plasma composition research has now matured to the point where the observed differences in behaviour between different element pairs are providing
further insights. 
Previous authors had noted that S sometimes behaved like a low-FIP element and other times like a high-FIP element but there was no obvious explanation
as to why. The development of the ponderomotive force model of the FIP effect \citep{Laming2004} led \cite{Laming2019} to argue that this behavior is driven
by the location of FIP fractionation and/or whether the magnetic field is open or closed.
Many interesting observational aspects of the 
FIP effect that set constraints for the models have therefore been found within recent studies. For example, it seems that the FIP effect \citep[detected using S
measurements from the Hinode/EUV Imaging Spectrometer (EIS),][]{Kosugi2007,Culhane2007} 
operates most
strongly at loop footpoints \citep{Baker2013,Baker2015}, which not only suggests that the loops are being filled by material from the top of the chromosphere, but is a fact
that can also be used as a tracer of SEP events \citep{Brooks2021b}.
Indications are that the IFIP effect is also strongest at the loop footpoints \citep{Brooks2018},
another constraint for models. 
To et al. (2024) and Brooks et al (2024) have both found differences in the magnitude of fractionation between Si/S and Ca/Ar ratios in a solar flare and post-X-flare
coronal rain observed by EIS. A possible explanation in both cases is that the energy deposition reaches the deep chromosphere, where S behaves like a
low-FIP element, and evaporates material from there.

\begin{figure*}
  \centerline{%
    \includegraphics[width=1.0\textwidth]{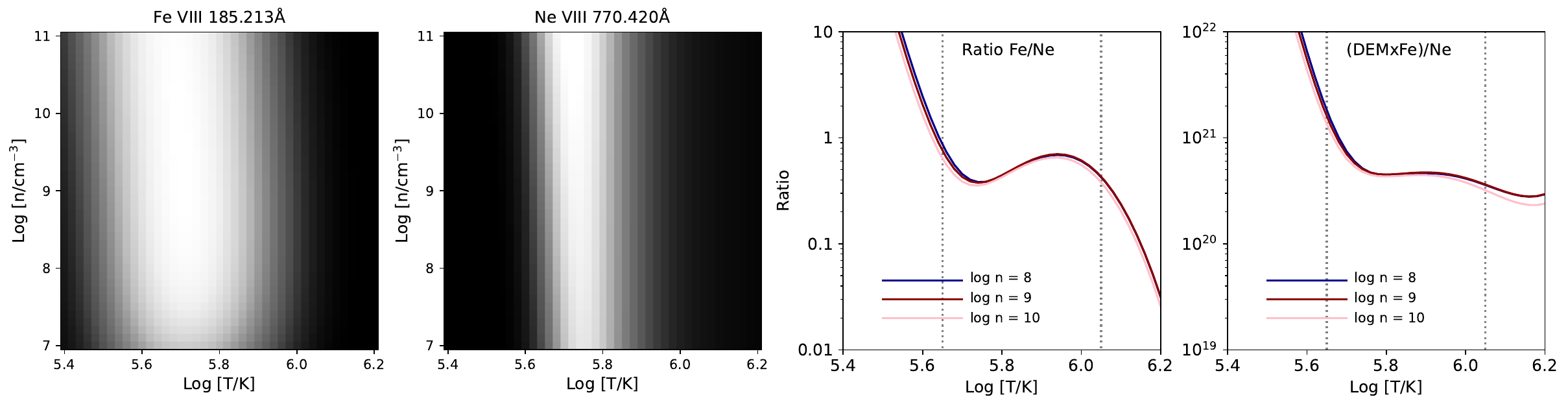}} %
  \caption{\ion{Fe}{8} 185.213\,\AA/\ion{Ne}{8} 770.428\,\AA\, abundance diagnostic ratio. 
First panel: \ion{Fe}{8} 185.213\,\AA\, contribution function plotted against temperature and density. 
Second panel: \ion{Ne}{8} 770.428\,\AA\, contribution function plotted against temperature and density. 
Third panel: ratio plotted against temperature for densities of $\log$ (n/cm$^{-3}$) = 8--10.
The ratio shows $\sim$40\% variation in the range of $\log$ (T/K) = 5.65-6.05 (delineated by the vertical
dotted lines). 
Fourth panel: ratio plotted against temperature for densities of $\log$ (n/cm$^{-3}$) = 8--10 after the \ion{Fe}{8} 185.213\,\AA\, contribution function has
been convolved with an idealised DEM (see text). Variations are reduced by a factor of 2 in the range of 
$\log$ (T/K) = 5.75-6.05.
}
  \label{fig1}
\end{figure*}

\cite{DelZanna2018} give a review of abundance measurements from EUV spectra and highlight some of the contradictory results in the literature.
Further investigations are needed and are critical for achieving the goals of the Solar Orbiter mission \citep{Muller2020}, which aims to match remote sensing abundance
measurements from SPICE \citep[][]{SPICE2020} with in-situ measurements of the solar wind detected by SWA \citep[Solar Wind Analyser,][]{Owen2020}, and the 
next generation Solar-C
EUVST mission \citep{Shimizu2020} that will have significantly improved sensitivity and spatial and temporal resolution compared to EIS. 
The observable wavelength range of SPICE makes
this a challenging task \citep{Brooks2022b}. 

A full understanding of the capabilities
and limitations of abundance diagnostics is key to the success of these missions and is now overdue.
Initial studies took advantage of the very clear separation in fractionation magnitude (FIP bias) from the photosphere to the corona, but much of the recent work 
has moved towards
looking at smaller scale spatial and temporal variations closer to the limits of the uncertainties, or, for example, the unusual behavior of S, which, as discussed,
potentially allows differentiation between fractionation at the top of the chromosphere from fractionation lower down. The degree of localization possible would
depend on the measurement precision, and verification of the accuracy, which is what we address here. 

In this article, we assess 
the \ion{Fe}{8} 185.213/\ion{Ne}{8} 770.428 ratio which is insensitive to logarithmic temperatures in the range of 5.65--6.05, and can be observed as a combination using
EIS and SPICE. In principle it can be
used to measure impulsive events, bright active region fan loops, coronal holes, and active region outflows, but we perform an empirical test 
to address the question of how accurately this abundance diagnostic can recover the FIP bias (magnitude of coronal to photospheric abundance)
from observed solar features with a known composition. 

By way of illustration, we apply the the diagnostic to coordinated observations from EIS and SPICE from the 2023 March Solar Orbiter perihelion, and measure
the EIS+SPICE composition. The diagnostic provides a new method for multi-mission
multi-viewpoint tracing of the solar wind to its solar sources.

\section{Fe/Ne elemental abundance diagnostic ratio}
One of the strongest lines observed by SPICE is \ion{Ne}{8} 770.428\,\AA. The theoretical contribution function (G(T,n) - where T is temperature and n is electron density) 
for this line peaks near $\log$ (T/K) = 5.75, and it overlaps
with emission from the strong \ion{Fe}{8} 185.213\,\AA\, line that is observable with EIS.
Figure \ref{fig1} shows the contribution functions for these two lines, and the theoretical ratio,
as a function of temperature and density. \ion{Fe}{8} 185.213\,\AA\, is broader than \ion{Ne}{8} 770.428\,\AA\, and peaks at $\log$ (T/K) = 5.7 but there is clear
overlap. \ion{Fe}{8} 185.213\,\AA\, is blended with \ion{Ni}{16} 185.23\,\AA\, \citep{Young2007}, which forms at a much higher temperature of $\log$ (T/K) = 6.45. 
\ion{Ni}{16} 185.23\,\AA\, therefore dominates the emission in the cores of active regions but has a negligible contribution
at lower temperatures \cite[see e.g. Figure 4 of ][]{Brooks2011b}.

The density dependence of the \ion{Fe}{8} 185.213/\ion{Ne}{8} 770.428 ratio is minimal in the $\log$ (n/cm$^{-3}$) = 8--10 range. The ratio also only varies by $\sim$ 40\% in the temperature range of $\log$ (T/K) = 5.65--6.05, although
it then deviates significantly at higher and lower temperatures. For a robust measurement then, the ratio should be convolved with the differential emission measure (DEM)
distribution of the structure of interest. This flattens and removes the temperature sensitivity as shown in the right panel in Figure \ref{fig1} where convolution
with a customized inverse Gaussian DEM (peak at $\log$ (T/K) = 6, width of $\log \sigma_T$ = 5.45) 
reduces the variation by a factor of 2 in the range of $\log$ (T/K) = 5.75--6.05. In Section \ref{bem} we derive real Sun DEM distributions to
perform a test of the accuracy of the diagnostic, but the ratio does appear to be potentially useful in a relatively unexplored temperature range. At least for
EIS there are no very good abundance diagnostics for the upper transition region \citep{Feldman2009}. There have, however, been some previous studies of this
region using SOHO \citep{Young1997}.

\begin{figure}[ht]
  \centerline{%
    \includegraphics[width=0.5\textwidth]{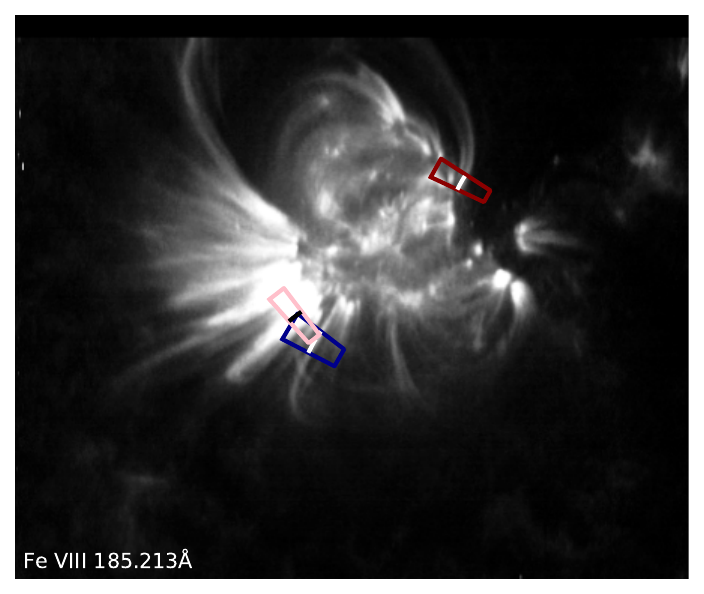}} %
  \caption{ Loops segments selected for the emission measure analysis. The \ion{Fe}{8} 185.213\,\AA\, image is from 
an EIS raster scan that started at 00:19:27\,UT\, on 2007 December 10. The loop segments are shown by the dark blue,
dark red, and pink colored boxes. These colors are also used in Figures \ref{fig3} and \ref{fig4} to cross-reference the
loops with their corresponding EM distributions and empirical test results.
}
  \label{fig2}
\end{figure}
Note that the G(T,n) function contains all of the relevant atomic physics parameters including spontaneous radiative decay rates, electron collisional excitation and 
deexcitation coefficients, ionization fractions, and elemental abundance. We calculated these functions using the CHIANTI v.10 database \citep{DelZanna2021} 
supplemented with density dependent 
ionization equilibrium calculations for the light ions from ADAS \citep{Summers2006}. For the prime ions \ion{Fe}{8} and \ion{Ne}{8} 
the radiative data and effective collision strengths were 
taken from \cite{DelZanna2014b}, \cite{Badnell2011}, and \cite{Liang2011}. We used the photospheric abundances of \cite{Scott2015a} and \cite{Scott2015b}.

\section{Baseline Emission Measure (EM) distributions for the empirical test}
\label{bem}
Our objective is to determine how accurately we can measure the FIP bias using the \ion{Fe}{8} 185.213/\ion{Ne}{8} 770.428 ratio. For our empirical test, we collect
a sample of observed intensities based on different (unknown) FIP bias magnitudes, and then attempt to recover these $``$ground truth$"$ values.
Prior to performing our empirical test, however, we require a sample of EM distributions as a baseline that can then be modified by the degree of FIP bias in the test
in order to compute the line intensities. In
addition we want to probe the accuracy of the ratio in different scenarios, and particularly when the target feature is formed close to, and outside, the region where the ratio is 
relatively flat in temperature to see what impact these differences have.
Therefore we derive the EM distributions for three fan loops observed by EIS. 

For this purpose we use an EIS dataset that was obtained on 2007 December 10 starting at 00:19:27\,UT. This observation used
the 1$''$ slit to scan AR 10978 obtaining a field-of-view (FOV) of 460$'' \times$384$''$ in 5 hours 19 mins. The exposure
time at each slit position was 40\,s. The study acronymn was AR\_velocity\_map\_v2.
This observation telemetered a very 
large spectral line-list to ground, and thus included 
lines from multiple ions of many species. 

The EIS data were processed using the standard (eis\_prep) routine available in SolarSoftware \citep{Freeland1998}.
After correcting the data for the effects of the dark current pedestal and contaminated pixels, the photometric
calibration is applied to convert the count rates to erg cm$^{-2}$ s$^{-1}$ sr$^{-1}$ \AA$^{-1}$. We used the most recent
updated radiometric calibration \citep{DelZanna2023}.

Figure \ref{fig2} shows AR 10978 and the loops we analyze. We selected many fan loops but these three are shown because they have contrasting EM distributions and
they peak at different temperatures. They therefore
probe the validity of the ratio in the regions we are seeking to test. The
examples are 
shown in the figure in the dark blue, dark red, and pink boxes. 
For measuring the loop intensities, we adopt the method of \cite{Aschwanden2008} as implemented by \cite{Warren2008}. In this method, we extract
the cross loop 
intensity profile at each position along the loop axis within 
a selected segment. These profiles are then averaged along 
the axis and the loop
is visually identified by selecting two background positions as close to it as possible. 
The averaged cross loop intensity profile is then fit with a Gaussian function and linear background. The background is subtracted and the area under the Gaussian curve is
then the loop intensity. 
This procedure is repeated for a range of emission lines covering a wide range of temperatures from
\ion{O}{4}--\ion{O}{6} (formation temperatures; $\log$ (T/K) = 5.2--5.5),
\ion{Mg}{7} ($\log$ (T/K) = 5.8), and
\ion{Fe}{8}--\ion{Fe}{13} ($\log$ (T/K) = 5.7--6.2).
Only the intensities, the cross-loop line profiles of which are highly correlated (linear Pearson coefficient $r>0.8$) with that of 
\ion{Fe}{8} 185.213\,\AA\, are included in the EM analysis. The other line intensities are set to zero and the uncertainty is estimated to be 20\% of the
background intensity. The uncertainties on the included line intensities come from the photometric calibration \citep[$\sim$23\%,][]{Lang2006}.
The G(T,n) functions for the EM calculation are computed assuming an initial density of $\log$ (n/cm$^{-3}$) = 9.3, which is typical for active region loops \citep{Brooks2012}.
The coronal abundances of \cite{Schmelz2012} were used for these calculations.

\begin{figure*}
  \centerline{%
    \includegraphics[width=1.0\textwidth]{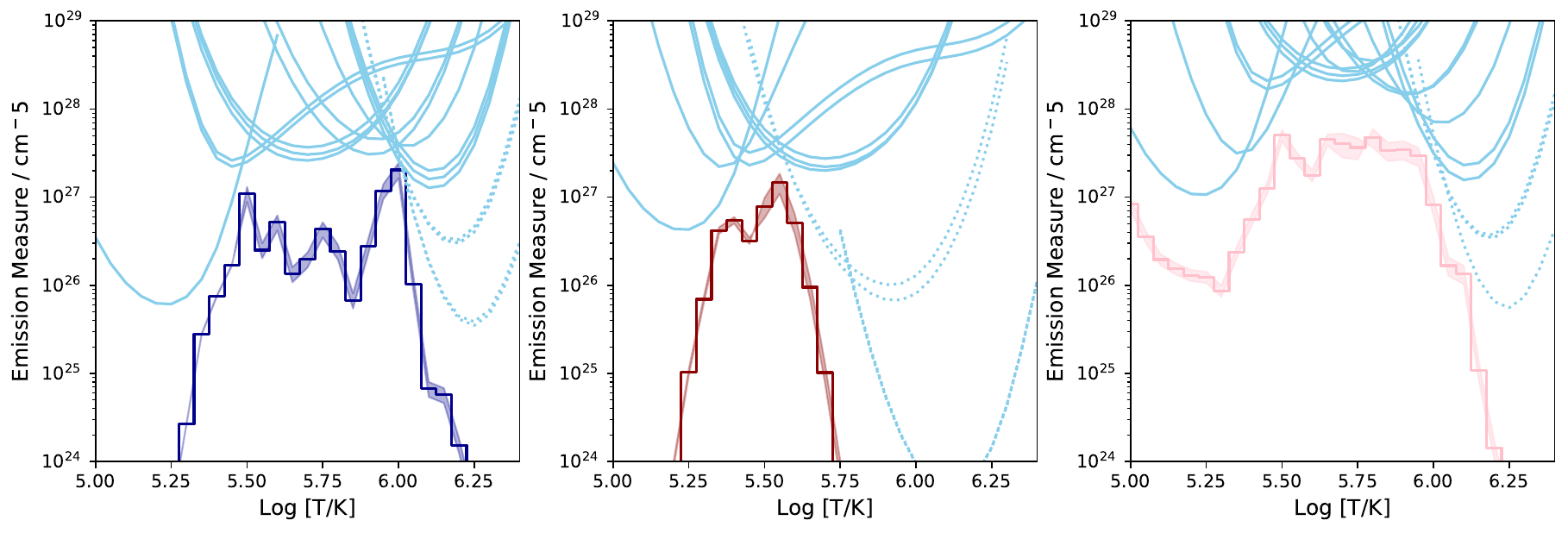}} %
  \caption{ Loop emission measure distributions. The histograms show the best-fit EM solutions from the MCMC algorithm.
They are color coded dark blue, dark red, and pink for cross-referencing with the selected segments shown
in Figure \ref{fig2} and the empirical test results of Figure \ref{fig4}. The light blue solid and dotted lines are EM loci
curves that constrain where the EM distribution can go. They represent emission lines from various ionization stages
of \ion{Fe}{8}--\ion{Fe}{13}, \ion{O}{4}--\ion{O}{6}, and \ion{Mg}{7}, and \ion{Si}{7}. The dotted curves indicate that
the cross-loop intensity profile is not well correlated with that of \ion{Fe}{8} 185.213\,\AA.
The shaded area around the EM distribution indicates the uncertainty, based on a comparison of the observed and predicted
intensities (see text for discussion).
}
  \label{fig3}
\end{figure*}
To derive the EM distributions we use the Markov Chain Monte Carlo (MCMC) method available in the PintOfAle software package \citep{Kashyap1998,Kashyap2000}. 
This algorithm reconstructs the EM by finding the best-fit solution that reproduces the observed intensities. The best-fit is obtained by perturbing the intensities
within the calibration uncertainty and running multiple reconstructions (100 in our case). Figure \ref{fig3} shows the best-fit solutions for the three fan loops in dark
blue, dark red, and pink (for cross referencing with the selected segments in Figure \ref{fig2}). The uncertainties in the solution, based on the differences between the
the observed and calculated intensities, are shown by the shaded areas. The shaded
areas fit tightly around the best-fit solutions. This indicates that they are reproducing the observed intensities well. In fact, for the three examples we show, all but the \ion{Mg}{7} 278.402\,\AA\, line 
are reproduced to within 34\%. 

The curves are EM loci, and indicate where the EM solutions are constrained. They basically represent the inverse
of the G(T,n) functions assuming that the observed intensity is zero. The solid loci are for lines included in the calculation. The dotted loci are for lines that were
excluded because their cross-loop intensity profiles were not highly correlated with those of \ion{Fe}{8} 185.213\,\AA. 
These are the baseline EM distributions that we deploy in the
empirical test. Note the complex structure in the dark blue distribution. This can arise from a number of uncertainties but
we retain it for our experiment since the unusual shape provides a further test of our methods.

\begin{figure*}
  \centerline{%
    \includegraphics[width=1.0\textwidth]{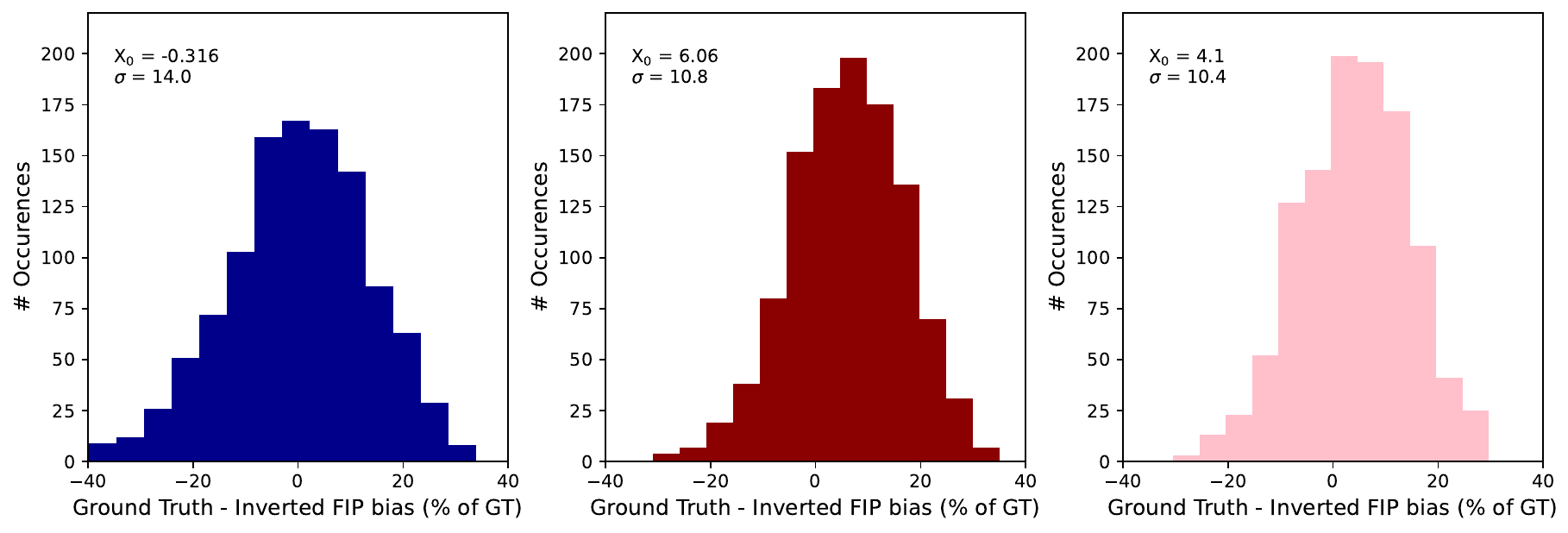}} %
  \caption{ Results from the empirical tests. The histograms show the difference between the ground truth and inverted
FIP bias as a percentage of the ground truth for 1100 experiments. They are color coded dark blue, dark red, and pink for cross-referencing with the selected segments shown
in Figure \ref{fig2} and their corresponding EM distributions shown in Figure \ref{fig3}. We also show the standard deviation ($\sigma$) and 
centroid of the peak of the distributions (X$_0$) in the legends.
}
  \label{fig4}
\end{figure*}

\section{Empirical Test}
\label{emptest}
The baseline EM distributions from the EIS observations are used to simulate the observed intensities for a range of FIP bias values. 
Since the EM distributions were derived assuming the coronal abundances of \cite{Schmelz2012} (a FIP bias of 2.4 for the Fe lines used in our tests), 
they are scaled by a factor of 0.4-1.7 to simulate a range of FIP bias magnitudes from a photospheric value of 1.0 to a strong 
FIP effect coronal value of 4.
The magnitude of the FIP bias is chosen randomly in the experiment. The EM is then 
combined with these FIP bias values to predict the observed intensities, and these are what are used for the inversion algorithm 
in the empitical test. 

In the new method we apply here, we use lines from \ion{Fe}{7}--\ion{Fe}{14}. These are essentially the same lines that we used for deriving the EIS EM distributions except with the
addition of an \ion{Fe}{7} 195.392\,\AA\, line identified by \cite{Young2021} and the \ion{Fe}{14} 211.316\,\AA\ line to provide extra low and high temperature
constraints. The full list is:
\ion{Fe}{7} 195.392\,\AA,
\ion{Fe}{8} 185.213\,\AA,
\ion{Fe}{9} 188.497\,\AA,
\ion{Fe}{10} 184.536\,\AA,
\ion{Fe}{11} 188.299\,\AA,
\ion{Fe}{12} 195.119\,\AA,
\ion{Fe}{13} 202.044\,\AA,
\ion{Fe}{13} 203.836\,\AA,
\ion{Fe}{14} 211.316\,\AA.
We use these lines so that the EM solution is derived abundance free. The method is the same as described above, it is just the spectral line list
that is adjusted. The FIP bias is determined from the ratio of the EM predicted to observed \ion{Ne}{8} 770.428\,\AA\, intensity.

To summarize the algorithm, the following steps are performed:
\begin{itemize}
\item EIS observed fan loop intensities $\rightarrow$ baseline DEM
\item DEM + random FIP bias $\rightarrow$ Fe and Ne line intensities
\item Fe and Ne line intensities $\rightarrow$ DEM + derived FIP bias
\end{itemize}

We show the results of this analysis in Figure \ref{fig4}, where histograms of the difference between the input ground truth FIP bias and the measurement from the inversion 
technique are plotted for the dark blue, dark red, and pink loop segments in the dark blue, dark red, and pink histograms. 
We find that, in general, the accuracy of the FIP bias - defined as the standard deviation of the histogram - is within 10.4-14.0\%
which is highly
encouraging. 
The peaks of the measurement distributions are also very close to the ground truth ($<$6\% difference) in all cases. We chose this sample of 
fan loops because their EM distributions probe the different temperature sensitivities of the diagnostic ratio, but there does not seem to be any obvious impact,
for example as a result of the EM peak temperature, 
on the accuracy of the measurements. The small deviations we find 
are within the measurement uncertainty in all three examples. 
Our results suggest that we can use this technique to measure the \ion{Fe}{8} 185.213/\ion{Ne}{8} 770.428 FIP bias ratio quite successfully.

\section{Coordinated observations from Hinode/EIS and Solar Orbiter/SPICE}
\label{spice}
Using simultaneous EIS and SPICE observations of the same target on 2023 March 30,
we demonstrate the practical use of the \ion{Fe}{8} 185.213\,\AA/\ion{Ne}{8} 770.428\,\AA\, diagnostic.
This was during the first of three Solar Orbiter remote sensing windows in the 
second quarter of 2023. At the time of the SPICE observations, the spacecraft was close to the Earth-Sun line and at a heliocentric distance of 0.376\,AU.

The EIS dataset was obtained on 2023 March 30 starting at 18:25:56\,UT.
In this case the 1$''$ slit was used to scan AR 13262 and a nearby coronal hole and obtain an FOV of 240$'' \times$512$''$ in 3 hours 12 mins. This
study used 2$''$ coarse raster steps and the exposure time at each slit positions was 100\,s. The study acronymn was HPW021VEL240x512v2\_b.
The data were processed as described in Section \ref{bem}.

\begin{figure*}
  \centerline{%
    \includegraphics[width=1.0\textwidth]{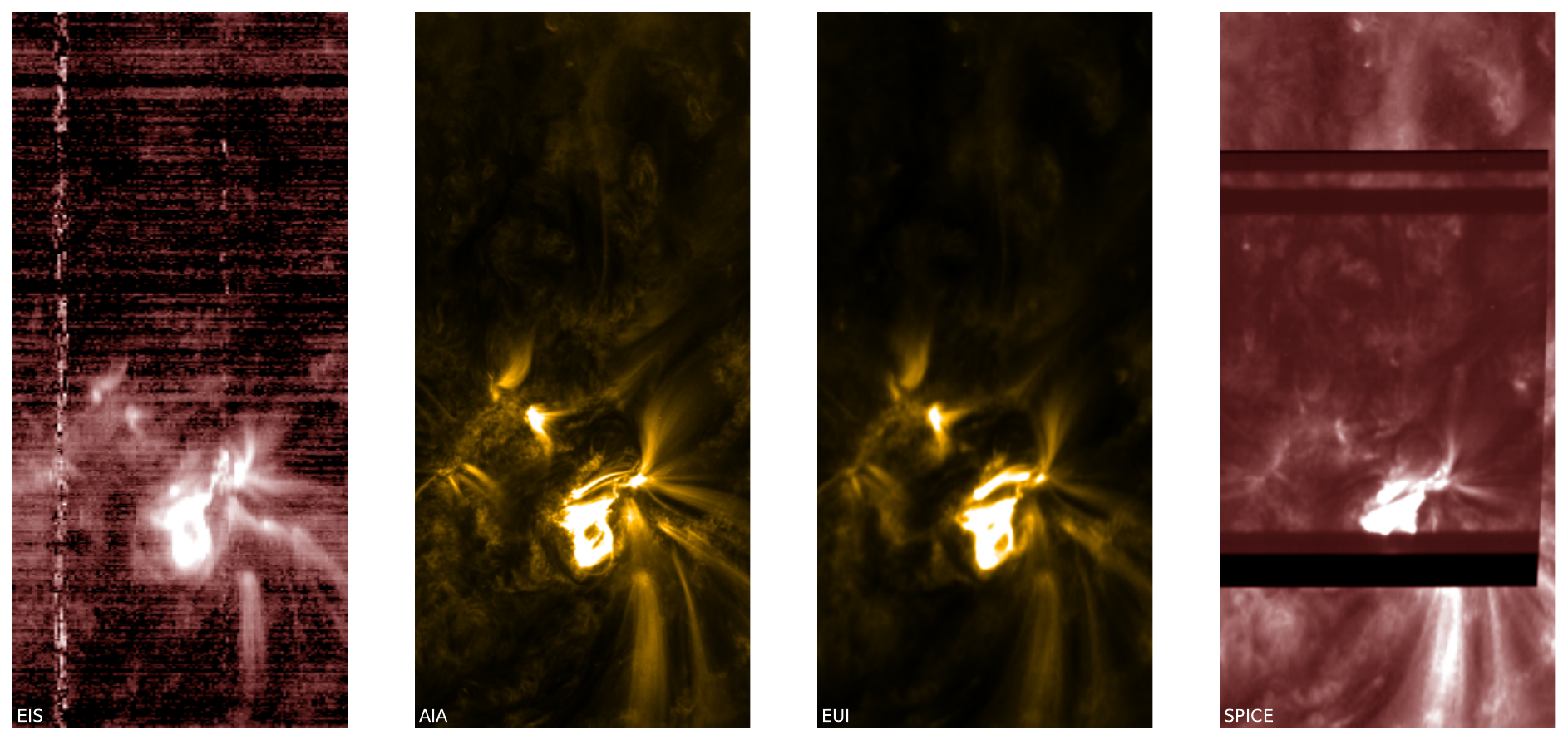}} %
  \caption{ Coordinated Hinode and Solar Orbiter observations from 2023 March 30. Left to right: EIS \ion{Fe}{8} 185.213\,\AA\, intensity from the raster
that started at 18:25:56\,UT, AIA 171\,\AA\, coaligned filter image taken at 20:10:57\,UT, EUI 174\,\AA\, coaligned filter image taken at 20:10:55\,UT,
SPICE \ion{Ne}{8} 770.428\,\AA\, coaligned waveband intensity from the raster scan that started at 19:29:28\,UT\, overlaid on the EUI image. The data
were coaligned as described in Section \ref{spice}.
}
  \label{fig5}
\end{figure*}
We use one SPICE dataset that was coordinated with the EIS observations and obtained on 2023 March 30 starting at 19:29:28\,UT.
This observation used the 4$''$ slit to scan an FOV of 768$'' \times$834$''$ in 1 hour 37 mins. The exposure time
at each slit position was 30\,s. The study acronymn was SCI\_DYN-DEEP\_SC\_SL04\_30.0S\_FF. The dataset includes nine
spectral line windows, but here we only use observations in \ion{Ne}{8} 770.428\,\AA.

The SPICE data were obtained from the Solar Orbiter level-2 data archive and were already calibrated to physical units of W m$^{-2}$ sr$^{-1}$ nm$^{-1}$
which we then converted to match the EIS data units. 
The estimated uncertainty in the measured intensities is $\sim$25\% \citep{SPICE2020}.
The SPICE observation ID is 184549538 with 002 as the unique identifier for the raster. The data we used for coalignment were from
SPICE Data Release 3.0 \citep{SPICEData3}, but our final results make use of re-calibrated data from SPICE Data Release 5.0 \citep{SPICEData5}. Note that
the updated data release quantitatively affects the measured \ion{Ne}{8} 770.428\,\AA\, intensities, but has no impact on the coalignment.

For context and coalignment, we also use data from the Atmospheric Imaging Assembly \citep[AIA,][]{Lemen2012} on the Solar
Dynamics Observatoiry \citep[SDO,][]{Pesnell2012} and the Extreme Ultraviolet Imager \citep[EUI,][]{Rochus2020} on Solar Orbiter.
The EUI/FSI (Full Sun Imager) image we use was also obtained from the Solar Orbiter level-2 archive and comes from EUI Data Release 6.0 \citep{EUIData6}.
The AIA data were obtained from the Joint Science Operations Center (JSOC) at Stanford University.

To compare the EIS and SPICE coordinated observations we coaligned the data from all four instruments. First, we coaligned the EIS \ion{Fe}{8} 185.213\,\AA\,
fitted intensity with the AIA 171\,\AA\, filter image. This was achieved by image cross-correlation to determine the EIS FOV in AIA coordinates and resample
the EIS data to the higher spatial resolution of AIA. We then coaligned the EUI 174\,\AA\, and AIA 171\,\AA\, full disk images. For this step, we created
map objects using the AstroPy \citep{astropy2022} and SunPy \citep{sunpy2020} packages and re-projected the EUI map to AIA Earth-view coordinates. Finally,
we used SunPy EUI and SPICE map objects to determine the SPICE FOV in EUI coordinates and cross-correlates the two images of the same FOV to accurately 
place the \ion{Ne}{8} 770.428\,\AA\, intensity image into the 
EUI map. This was then re-projected to the AIA Earth-view coordinates. In principle, the images from all four instruments are then coaligned. 

During re-projection, the EUI image is geometrically transformed to match AIA. Since the SPICE intensity image is placed in the EUI map, we assume that the 
transformation is the same for both instruments, however, the SPICE data are built up over more than an hour by a changing slit position, whereas the EUI image
is taken on a time-scale of seconds. Therefore, there are some inaccuracies in the re-projection for SPICE. In a final step, therefore, we used cross-correlation to improve 
the alignment between the EIS \ion{Fe}{8} 185.213\,\AA\, and SPICE \ion{Ne}{8} 770.428\,\AA\, images.

The observations were taken on 2023 March 30 as part of the Solar Orbiter Observing Plan (SOOP) (L\_SMALL\_HRES\_HCAD\_Slow-Wind-Connection) that is designed to ascertain
the sources of the slow solar wind. The coordinated campaign was designated as IRIS-Hinode Operation Plan (IHOP) 455
- Release of the slow solar wind at active region or coronal hole boundaries. A general overview of the goals of this SOOP are given in \cite{Zouganelis2020}, and a very detailed report
of the first campaign in 2022 March is provided by \cite{Yardley2023}. At the time of the SPICE and EIS observations, both instruments were observing 
the southern boundary of an equatorial coronal hole bordering AR 13262. This was the predicted location of the magnetic field connected to Solar Orbiter.
Figure \ref{fig5} shows the target region as observed in EIS \ion{Fe}{8} 185.213\,\AA, AIA 171\,\AA, EUI 174\,\AA, and SPICE  \ion{Ne}{8} 770.428\,\AA.

\begin{figure}[ht]
  \centerline{%
    \includegraphics[width=0.4\textwidth]{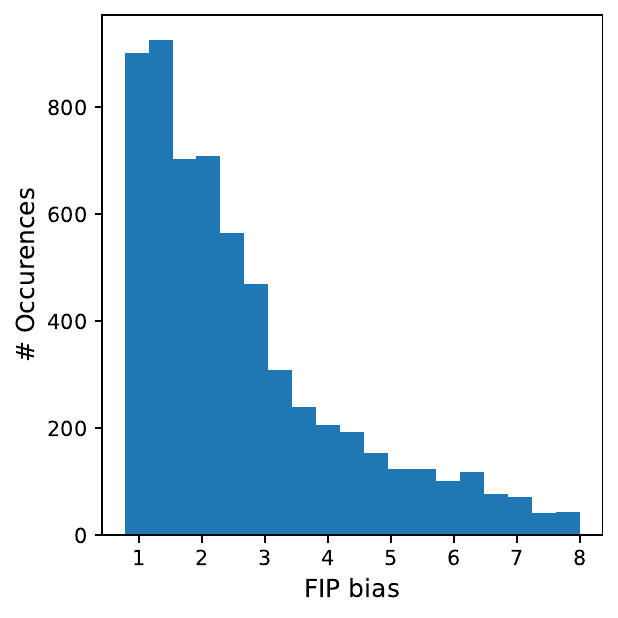}} %
  \caption{ Plasma composition analysis from the combined EIS and SPICE observations. The histogram shows the FIP bias measurements for the coaligned FOV shown in Figure \ref{fig7}.
}
  \label{fig6}
\end{figure}
We apply the technique developed in Section \ref{emptest} to the EIS and SPICE observations with one modification based on the available data: we use the  
\ion{Fe}{11} (188.216+188.219)/182.167 intensity ratio for the density diagnostics. The SPICE observed and EIS simulated \ion{Ne}{8} 770.428\,\AA\, intensities
are obtained at each coaligned pixel within the SPICE FOV. The ratio of these values is the FIP bias and we
show a histogram of the results for the coaligned EIS-SPICE FOV in Figure \ref{fig6}. 
They are filtered to remove uncertain results. The technique uses multiple lines of varying strengths so bad 
pixels or missing data or low signal-to-noise can have various effects. We removed pixels with non-finite or negative results, densities that fall outside the range
of the diagnostic ratio, $\chi^2$ values indicating a poor EIS EM solution, SPICE observed intensities with unreasonably large fitting errors, and FIP 
bias values outside the expected range of validity. The histogram peak is close to photospheric abundances. To compare to the in-situ data, however,
we need to know where the footpoints of the magnetic field connected to Solar Orbiter are located.

\begin{figure*}
  \centerline{%
    \includegraphics[width=1.0\textwidth]{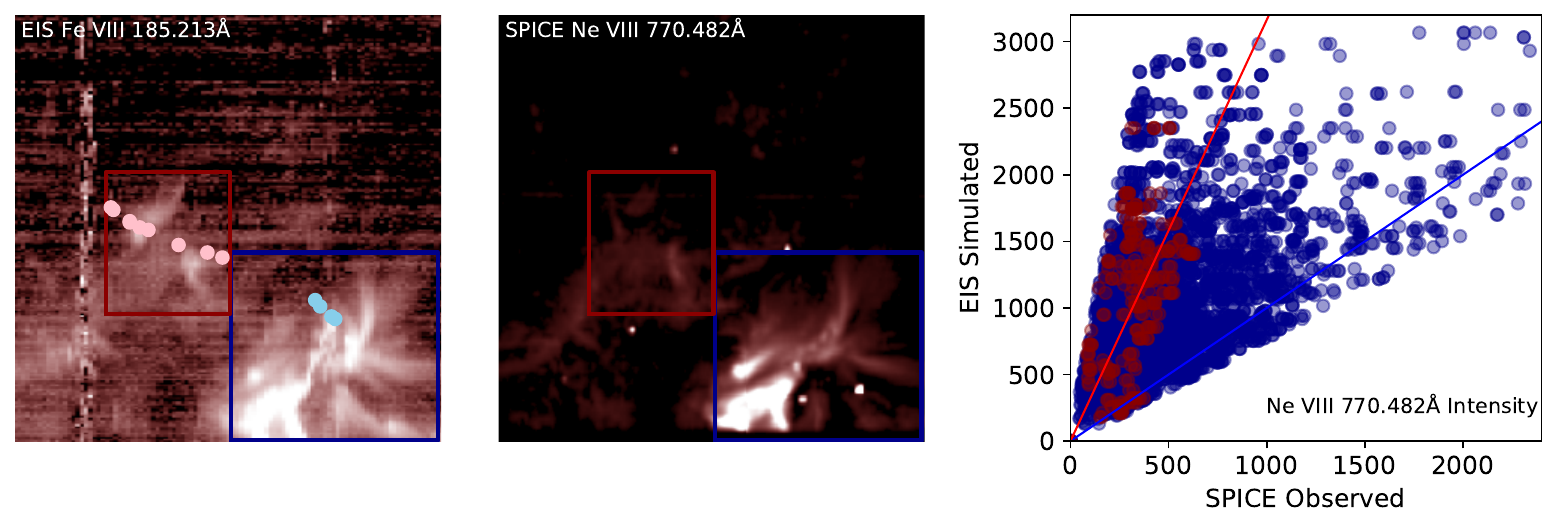}} %
  \caption{ Plasma composition analysis for the predicted footpoints of the magnetic field connected to Solar Orbiter. Left to right: EIS \ion{Fe}{8} 185.213\,\AA\, intensity map,
SPICE observed \ion{Ne}{8} 770.428\,\AA\ intensity map, EIS simulated and SPICE observed \ion{Ne}{8} 770.428\,\AA\ intensities for the blue and red
boxed regions in the left and middle panels. The red and blue boxes encompass regions where the magnetic field connected to Solar Orbiter is predicted
to be located. The pale blue and pink dots show the actual predictions for the footpoints of the magnetic field connected to the spacecraft (see text). The blue line in the right hand panel indicates photospheric abundance ratios, and the red line shows coronal abundance ratios.
}
  \label{fig7}
\end{figure*}

We therefore modeled the locations of the footpoints of the magnetic field connected to Solar Orbiter. We
back map the magnetic field from the spacecraft to the solar surface by using a Parker
spiral model to trace the heliospheric magnetic field to the source surface of a potential field extrapolation (PFSS). The shape of the Parker spiral is determined by the measured solar wind
speed from the SWA Proton and Alpha particle Sensor (PAS). The 
PFSS model is then used to trace the field from the source surface (in this case at 2.5\,R$_{\odot}$) to the coronal footpoints. We use the PFSS model in
the pfsspy python package \citep{Stansby2020}. The coronal magnetic field
is reconstructed from an ADAPT-GONG (Air force Data Assimilative Photospheric flux Transport) synoptic map \citep{Hickmann2015}. We used a map taken on 2023 March 30 at 20:00\,UT\, to match the EIS/SPICE timing.
The locations of the footpoint source regions predicted by the model are shown by the pale blue and pink dots on the EIS image in Figure \ref{fig7}. 
There is some discrepancy, however, between the model and the in-situ Solar
Orbiter wind speed data. 
The model finds that the footpoints fall within the blue box in the figure (covering the main leading polarity) until approximately the end of March 28 (estimated Sun time). They then migrate to the red box (covering decayed magnetic flux), before the connectivity later moves
off to the East. Within the time-period of the EIS observations the predicted connectivity is tracking across the red box. The in-situ
wind speed data, however, suggest that for some periods Solar Orbiter may still be sampling solar wind from the main polarity (blue box). Modeling the connectivity of course has
limitations due to the assumptions inherent in the methods.

\begin{figure*}
  \centerline{%
    \includegraphics[width=1.0\textwidth]{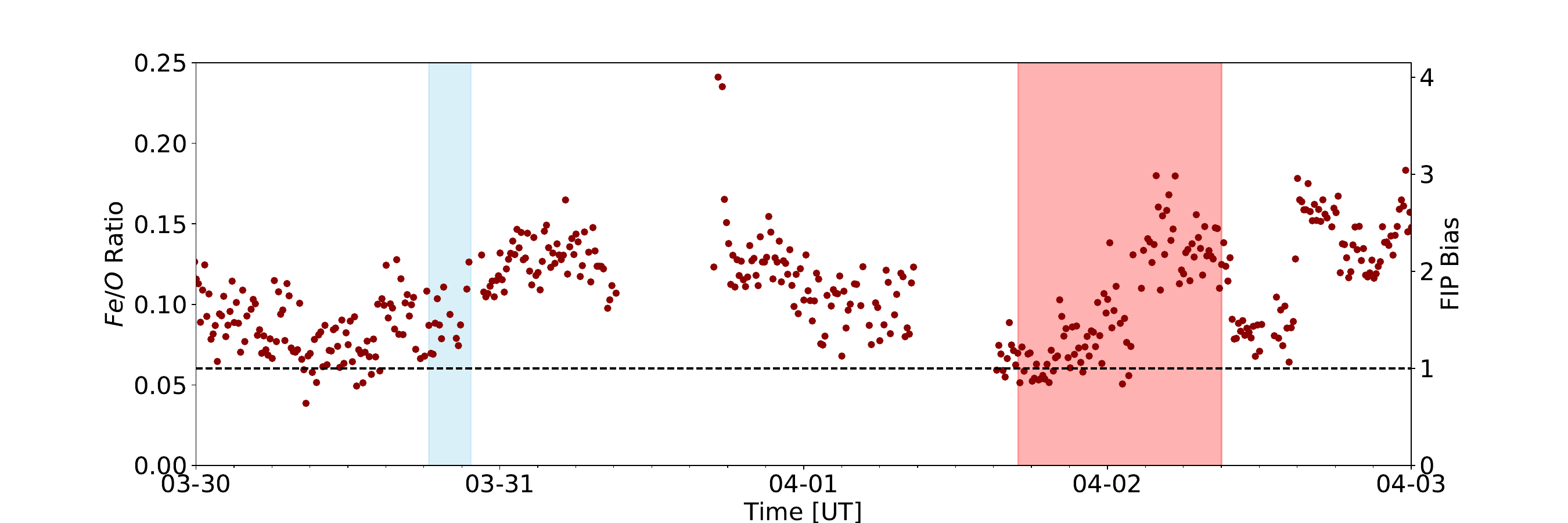}} %
  \caption{ SWA/HIS measurements of the Fe/O intensity ratio in-situ in the solar wind for the period 2023 March 30 to April 3. The corresponding FIP bias
values (Fe/O normalized to the photospheric ratio) are shown on the right axis. The horizontal dashed line indicates the photospheric ratio. The light blue
shaded area shows when EIS and SPICE were observing. The red shaded area indicates when solar wind from the AR 13262 arrives at Solar Orbiter.
}
  \label{fig8}
\end{figure*}

In the right hand panel we compare the SPICE observed and EIS simulated \ion{Ne}{8} 770.428\,\AA\, intensities for the two boxes. When the simulated and observed intensities match, the FIP bias is 1. This is indicated by the blue line in Figure \ref{fig7}. The red line indicates coronal abundances (FIP bias $\sim$ 3.2). There is a large variation from photospheric
to coronal abundances in the blue box. The measurements in the red box are much closer to the coronal abundances.
The results then indicate an increase in FIP bias on average as connectivity moves from the blue to the red box, with an initial contribution from photospheric abundances that is then dominated by coronal abundances. This is broadly in agreement with the SWA/Heavy Ion Sensor (HIS) measurements made
in-situ. Figure \ref{fig8} shows the HIS data for the period 2023 March 30 to April 3. The shaded red box shows the period when the solar wind arriving at Solar Orbiter appears to be coming
from AR 13262. The HIS
measurements do show an increase in the Fe/O ratio (FIP bias increase of a factor of $\sim$ 2.5 from photospheric to coronal) as the connectivity changes from the leading to the following polarity.
Note that we do not expect exact agreement between Fe/Ne and Fe/O. There is some evidence that O fractionates more than Ne in the corona, and also that the Ne/O ratio
varies with the solar cycle in the slow speed solar wind \citep{Shearer2014} and the Sun-as-a-star \citep{Brooks2018b}.

Note also that although the increasing trend in the Fe/Ne FIP bias from the blue box to the red box is unaffected, the absolute magnitude of the increase would be affected 
by changes in the SPICE photometric calibration. Recent work by \cite{Varesano2024} shows a systematic discrepancy between the quiet Sun intensities measured by SPICE and SOHO/SUMER
(see their Table 3). We have independently verified this by comparing SPICE quiet Sun intensities from \cite{Brooks2022b} with SUMER quiet Sun intensities from \cite{Warren2005}. 
It is not clear if this is a result of calibration uncertainties, or real changes on the Sun.

\section{Future observations by Solar-C/EUVST}
Accurate abundance diagnostics are highly valuable but challenging to find.
\cite{Feldman2009} discussed potential abundance diagnostics that fall within the EIS wavelength range and following his work the \ion{Si}{10} 258.375/\ion{S}{10} 264.223 
ratio has been widely used. Much of the literature reviewed in the introduction was built on this ratio, however, it forms at too high a temperature to overlap with
emission lines observed by SPICE. 
The goal of this work was to find an abundance diagnostic that could be observed using EIS and SPICE. The \ion{Fe}{8} 185.213/\ion{Ne}{8} 770.428 ratio appears to be useful.

The new method we apply here was focused on EIS and SPICE. Although coordination and coalignment of observations from multiple instruments is difficult, there is a growing
number of joint EIS and SPICE observations where the diagnostic can be used. Furthermore, this
method is applicable also to future observations by the Solar-C Extreme UltraViolet high-throughput
Spectroscopic Telescope \citep[EUVST,][]{Shimizu2020}. EUVST is a combined spectrograph and slit-jaw imager that will provide comprehensive
temperature coverage throughout the solar atmosphere (0.02--20\,MK) at high spatial (0.4$''$) and temporal (0.5\,s) resolution. The current design wavebands
cover 170--212.3\,\AA, 719.1--846.8\,\AA, 
927.6\,\AA, and 1115.0--1220.9\,\AA\, (Imada et al. 2024). 
EUVST will therefore observe spectral lines from Fe ionization stages up to \ion{Fe}{14} in 
non-flaring conditions. In fact, the spectral lines we used in our empirical test in Section \ref{emptest} were deliberately chosen to be observable by
EUVST. 

Although these wavebands do include abundance diagnostic ratios for flares,
such as \ion{Ca}{14} 193.874/\ion{Ar}{14} 194.396, they are relatively sparse in lines from high-FIP elements that form in the upper transition region or low corona.
The ratio we have examined is one good example that will be observed, however. 
The \ion{Fe}{8} 185.213\,\AA\ line falls in the short wavelength detector range, and \ion{Ne}{8} 770.428\,\AA\ falls in long-wavelength channel 1.
Based on our test results, this will open the possibility of accurate (15\%) FIP bias measurements of upper transition region structures at higher spatial and temporal
resolution than ever before.

\section{Conclusions}
We have demonstrated the effectiveness of the \ion{Fe}{8} 185.213/\ion{Ne}{8} 770.428 abundance diagnostic ratio for 
multi-spacecraft campaigns with Hinode/EIS and Solar Orbiter/SPICE. The ratio is most useful in the temperature range of $\log$ (T/K) = 5.65--6.05, but this can
be extended with appropriate consideration of temperature effects. We reiterate that \ion{Ni}{16} 185.23\,\AA\, largely dominates the emission in 
\ion{Fe}{8} 185.213\,\AA\, at higher temperatures ($\log$ (T/K) = 6.45), so the ratio is less useful in the hot cores of active regions.

Applying the diagnostic to an empirical test of FIP bias samples generated from real active region fan loop observations from EIS, we find that the measured FIP bias
is accurate to within 10--14\%. This is encouraging since it is well below the large range in composition variations from sub-photospheric (FIP bias below 1) to coronal (FIP bias of $\sim$3+) abundances that are observed in the 
solar atmosphere. It will be useful for research into such variations, for solar wind connection studies, and potentially also for stellar observations.

Analysis of coordinated EIS and SPICE observations from 2023 March demonstrates the value of the diagnostic for these campaigns. We were able to determine the Fe/Ne FIP bias
in the regions where the footpoints of the magnetic field connected to Solar Orbiter were predicted to be located. They appear to first be rooted in the leading polarity of
the active region and then to migrate across to decayed trailing polarity to the East. A spread of values from photospheric to coronal abundances in the leading polarity
increases on average to a coronal abundance in the trailing polarity. 
SWA sensor measurements taken in-situ in the spacecraft environment are broadly in agreement with the
remote sensing observations. 
The effectiveness of the diagnostic demonstrated here also bodes well for future studies using Solar-C/EUVST.

\begin{acknowledgments}
We thank the referee for helpful comments that improved the manuscript. We also thank Terje Fredvik for advice on the ongoing SPICE calibration activities.
The work of D.H.B. and H.P.W. was funded by the NASA Hinode program. 
D.B. is funded under Solar Orbiter EUI Operations grant number ST/X002012/1 and Hinode Ops Continuation 2022-25 grant number ST/X002063/1.
S.L.Y. is grateful to the Science Technology and Facilities Council for the award of an Ernest Rutherford Fellowship (ST/X003787/1).
Hinode is a Japanese mission developed and launched by ISAS/JAXA, collaborating with NAOJ as a domestic partner, NASA and STFC (UK) as international partners. Scientific operation of the Hinode mission is conducted by the Hinode science team organized at ISAS/JAXA. This team mainly consists of scientists from institutes in the partner countries. Support for the post-launch operation is provided by JAXA and NAOJ(Japan), STFC (U.K.), NASA, ESA, and NSC (Norway).
Solar Orbiter is a mission of international cooperation between ESA and NASA, operated by ESA.
The EUI instrument was built by CSL, IAS, MPS, MSSL/UCL, PMOD/WRC, ROB, LCF/IO with funding from the Belgian Federal Science Policy Office (BELSPO/PRODEX); the Centre National d’Etudes Spatiales (CNES); the UK Space Agency (UKSA); the Bundesministerium für Wirtschaft und Energie (BMWi) through the Deutsches Zentrum für Luft- und Raumfahrt (DLR); and the Swiss Space Office (SSO).
The development of SPICE has been funded by ESA member states and ESA. It was built and is operated by a multinational consortium of research institutes supported by their respective funding agencies: STFC RAL (UKSA, hard- ware lead), IAS (CNES, operations lead), GSFC (NASA), MPS (DLR), PMOD/WRC (Swiss Space Office), SwRI (NASA), and UiO (Norwegian Space Agency).
The AIA data are courtesy of NASA/SDO and the AIA, EVE, and HMI science teams.
CHIANTI is a collaborative project involving George Mason University, the University of Michigan (USA), University of Cambridge (UK) and NASA Goddard Space Flight Center (USA).

\end{acknowledgments}

\facilities{\emph{Hinode}/EIS, SDO/AIA, Solar Orbiter/EUI, Solar Orbiter/SPICE}

\end{document}